\documentclass[aps,pre,showpacs,twocolumn]{revtex4-1}

\usepackage{graphicx}
\usepackage{dcolumn}
\usepackage{bm}
\usepackage[utf8]{inputenc}
\usepackage[T1]{fontenc}
\usepackage{mathptmx}
\usepackage{etoolbox}
\usepackage{CJK}

\usepackage{ulem}
\usepackage{amssymb}
\usepackage{epsfig}
\usepackage{subfigure}
\usepackage{mathrsfs}
\usepackage{verbatim}
\usepackage{rotating}
\usepackage{longtable}
\usepackage{amsmath}
\usepackage{array,color}

\begin{document}
\title{Amplifier scheme: driven by direct-drive under $\sim$ 10 MJ laser toward inertial fusion energy}
\begin{CJK*}{GB}{gbsn}

\author{Ke Lan (À¶¿É)}  \email{lan_ke@iapcm.ac.cn}
\author{Xiumei Qiao (ÇÇÐã÷)}
\author{Yongsheng Li (ÀîÓÀÉý)}
\affiliation{Institute of Applied Physics and Computational Mathematics, Beijing 100094, China}
\author{Xiaohui Zhao (ÕÔÏþêÍ)}
\author{Zhan Sui (ËåÕ¹)}
\affiliation{Shanghai Institute of Laser Plasma, China Academy of Engineering Physics, Shanghai 201800, China}


\begin{abstract}
The National Ignition Facility successfully achieved target gain 2.4 thus marginally entering into
burn stage. Meanwhile, a recent conceptual design on 10 MJ laser driver [{\it Matter Radiat. Extremes} {\bf 9}, 043002 (2024)] provides a new room for exploring novel target designs and interesting phenomena in a burning plasma after ignition.
In this paper, we propose an amplifier scheme with extended burn stage, which includes secondary implosion, generates extremely hot and dense fusion fireball, and produces additional gain.
Same as the central ignition scheme used for inertial confinement fusion since 1960s',
this new amplifier scheme includes implosion and stagnation, with fusion starting from the central hot spot and serving as a spark plug for ignition. However, the fuel burn in the amplifier scheme is dominated by density and has following characteristics: (1) formation of an extremely compressed shell with a high areal density at stagnation; (2) density dominated ignition and move of fusion peak toward the shell; (3) primary explosion in shell and formation of a fireball in the center; (4) secondary explosion in the extremely hot and dense fireball. The amplifier scheme can be realized either by direct-drive or by indirect-drive. Here, we present a direct-drive amplifier design. A central ignition design is also presented for comparison.
From our 1D simulations,
the yield released by the amplifier capsule after bangtime is 4.8 times that before, remarkably higher than 1.25 times of the central ignition capsule.
The amplifier scheme can be realized at a low convergence ratio, so it can greatly relax the $\rho R T$ hot spot condition and the stringent requirements on  engineering issues by a high gain fusion.
Especially, the fireball  lasts for 30 ps, reaching 330 g/cm$^3$, 350 keV, 54 Tbar at center when the secondary explosion happens, which leaves an important
room for novel target designs towards clean fusion energy.
\end{abstract}

\maketitle

Fusion energy has been a quest of mankind for more than a half century \cite{MTV, IFEneedsreport}. Ignition of the National Ignition Facility (NIF) at Lawrence Livermore National Laboratory \cite{Abu-Shawareb2024PRL, Hurricane2024PRL, Rubery2024PRL, Pak2024PRE, Kritcher2024PRE}
successfully demonstrated the feasibility of  inertial confinement fusion (ICF), which  ideally positions inertial fusion energy  (IFE) as a highly promising approach for energy production.
Current record   of the target energy gain is G = 2.4  on the NIF\cite{5.2MJ}, where G is the   ratio of  fusion energy output to  input laser energy, is far below   IFE requirements.
While G $>$1 implies that some fuel is burnt, its amount is small and the burn stage is not well understood.
In fact, burn efficiency $\Phi$ is a key to increase G. For IFE required high gain, one has to achieve extreme levels of spherical compression of fuel to reach high areal density $\rho R$  and high ion temperature  $T_i$ for sufficient burn-up. Here,  $\rho$ is mass density, and $R$ is radius. For fuel of hydrogen isotopes deuterium and tritium  (indicated as symbols D and T, respectively) mainly used in current ICF studies, it requires $\rho R$ $\sim$ 3 g/cm$^2$ and $T_i >$  4.3 keV with a density compression of about 1500 or a size convergence ratio of about 35. While, in the central ignition scheme used for ICF since 1960s', the ignition happens in a small central hot spot  fuel  with   mass  only a few percent of total fuel, and it is very hard to achieve such high level spherical compression by considering the drive asymmetry \cite{Lan2014POP, Divol2017POP, Craxton2020DPP, Lan2021PRL, Lan2022MRE, Ralph2024NP},
laser-plasma instabilities \cite{Strozzi2017PRL, Tikhonchuk2021MRE, incoherent2023MRE}, hydrodynamic instabilities \cite{Goncharov2000POP, Clark2018POP, Qiao2021PRL, Do2022PRL} and many engineering factors. As a result, it is hard to achieve a burn efficiency   higher than 30$\%$ in the central ignition scheme of ICF.

Nevertheless, it is interesting to note that the volumetric reaction rate is proportional to $\rho^2$, indicating the role of $\rho$ of the fuel in achieving efficient release of fusion energy\cite{MTV}.
A recent conceptual design of a 10 MJ laser driver \cite{10MJ} provides a  room for exploring novel target designs and interesting phenomena in burn stage to achieve   high burn efficiency  and  high gain.
In this paper, we will take this advantage of  $\rho$ and propose an amplifier scheme, which can generate an extremely hot and dense fusion fireball and produce additional gain after ignition via cascading explosions.
This amplifier scheme can be realized either by direct-drive or by indirect-drive.
Here, we will present  an amplifier design with a spherical capsule containing DT fuel inside an ablator shell of glow discharge polymer plastic (CH) for  direct-drive ICF \cite{Nuckolls1972Nature, Campbell2017MRE, Gopalaswamy2019Nature}. An indirect-driven amplifier design is presented in our separate paper \cite{LYS2024POP}.

\begin{figure*}[t]
\includegraphics[width=0.9\textwidth]{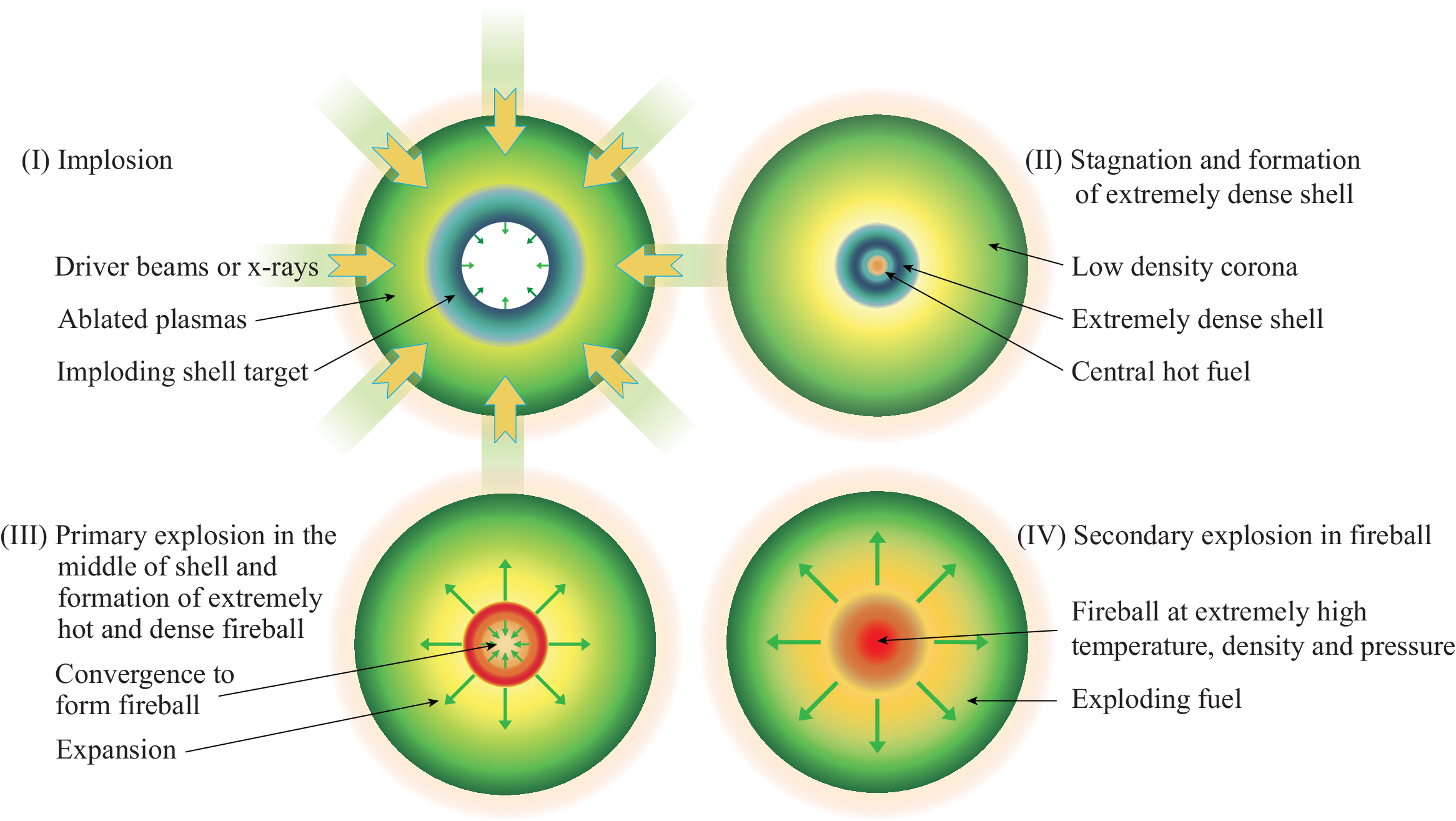}
\caption{Principle of the amplifier scheme. Here, we take the style of Fig. 3.1 in Ref. \citenum{MTV}}
\label{Fig:1}
\end{figure*}

Before presenting the  amplifier scheme, it is worth  mentioning the stages of the central ignition scheme  \cite{MTV, Lindl2018POP}, including:
(1) ablation and implosion;
(2) stagnation and formation of a hot spot;
(3) ignition in the hot spot;
(4) burn and explosion.
At stagnation, it generates a highly compressed fuel shell enclosing a hot spot in the center  with  $\rho_c/\rho_h$ $\sim$ 10, where $\rho_c$ and $\rho_h$ are densities of cold shell and hot spot respectively.
This scheme has following characteristics:
(1) pressure $P$ in  hot spot is constant, and both temperature and fusion reaction rate drop  sharply when approaching  the shell;
(2) it has an explosion happening in the central hot spot;
(3) whole fuel expands immediately after  explosion.
The characteristic  times include:
(1) $t_{fuel}$, when the released fusion energy is equal the thermal energy deposited in the hot spot, G$_{fuel}$ = 1;
(2) $t_{stag}$, stagnation time when kinetic energy of fuel in the shell attains its minimum;
(3) $t_{ign}$, ignition time when fusion
energy output equals to input laser energy, G $=$ 1;
(4) $t_{bang}$, bang time when the nuclear reaction rate $dN/dt$ of whole fuel reaches its peak.
Here, $N$ is number of neutrons, and $t$ is time.

Same as the central ignition scheme, the amplifier scheme includes  implosion, stagnation and ignition, with  fusion starting from the central hot spot and serving as a spark plug for ignition.
However, the amplifier scheme also includes a phase of secondary explosion and has following characteristics.
(1) {\it Formation of an extremely compressed shell with a high areal density at stagnation}.
(2) {\it Density dominated ignition and move of fusion peak from central region toward the shell}.
In this phase, the peaks of  pressure and $\frac{dN}{dmdt}$ are dominated by density and move rapidly toward the shell.
Here,  $m$ is mass, and  $\frac{dN}{dmdt}$ is reaction rate   per unit mass.
(3) {\it  Primary explosion in shell and formation of a fireball in the center}. When the peak of pressure moves  to the middle of shell  where density peaks, the primary explosion happens inside the extremely dense shell.  The primary explosion in shell pushes the inner part of fuel to converge spherically and meanwhile forms a fireball  in the center.
(4) {\it Secondary explosion and expansion}. When the inner shock converges at center,   the secondary explosion  happens inside the extremely hot and dense fuel and produces  supplementary fusion gain.
Principle of the amplified scheme is given in  Fig. \ref{Fig:1}.

In addition to  $t_{fuel}$, $t_{stag}$ and  $t_{ign}$, the amplifier scheme also has characteristic  times of $t_{pri}$ and $t_{sec}$.
Here, $t_{pri}$ is the primary explosion  time when $\frac{dN}{dmdt}$ reaches  peak in the extremely dense shell, and $t_{sec}$ is the secondary explosion time when $\frac{dN}{dmdt}$ reaches  peak at the fuel center.
The nuclear reaction rate of whole fuel can reach  its peak either at $t_{pri}$ or at $t_{sec}$,  depending on design.

\begin{figure*}[t]
\centering
\includegraphics[width=1.0\textwidth]{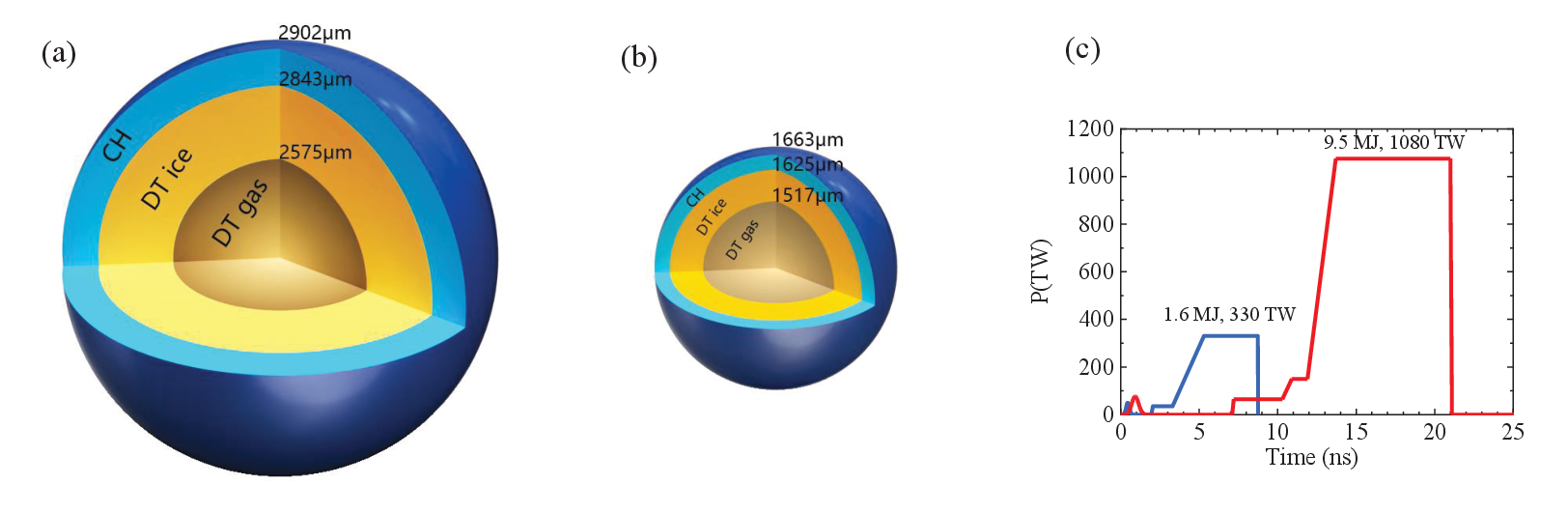}
\caption{\label{fig:sketch}(Color online) Schematics of the amplifier capsule  (a),   the central ignition capsule (b), and their drive laser profiles (c).}
\label{Fig:2}
\end{figure*}

In our direct-driven amplifier design, we consider a 2902-$\mu$m-radius spherical cryogenic capsule driven by a 9.46  MJ laser peaking at 1080 TW.
The capsule  contains a 6.33 mg DT fuel layer  and a 7.02 mg CH outer ablator layer.
To compare the different characteristics between the amplifier scheme and the central ignition scheme,
we also present a  central ignition design, which is  a 1663-$\mu$m-radius cryogenic capsule with 0.842 mg DT fuel and 1.38 mg CH ablator  driven by a 1.6 MJ laser peaking at maximum power of 330 TW. The schematics of two capsules and their drive laser pulses  are shown in Fig.\ref{Fig:2}.

Simulations studying  one-dimensional (1D) implosion performances  were performed with a multi-group radiation hydrodynamic code RDMG,  widely used in ICF studies \cite{Feng1999, Lan2017POP, Qiao2019PPCF}.
In RDMG, the  change rate of $E$, the internal energy per unit mass of fuel,  is written as:
\begin{eqnarray}
\frac{dE}{dt} = W_{dep} +  W_{m} +  W_{r} +  W_{i} +  W_{e} \nonumber
\end{eqnarray}
Here,  $W_{dep}$ is the power density deposited by  fusion products,  $W_m$ is the contribution due to mechanical work, $W_r$ is   energy lost rate by radiation via bremsstrahlung, and $W_e$ and  $W_i$ are   lost rates  by thermal conduction of electron  and   ion, respectively.
In direct-drive,  laser energy can be lost in the following ways. First, part of the laser beams  can miss   target as the capsule implodes. Second, the laser beams can be refracted in the plasma corona if they are  not normal to the capsule surface.
Third, the laser absorption  can be degraded by cross-beam energy transfer, which can divert energy away from the incoming laser beam into the outgoing rays.
In RDMG,  only the first two effects are considered and  described by means of a 3D ray-tracing algorithm with inverse bremsstrahlung absorption as the main absorption scheme.
We take the initial diameter of laser beam as $90\%$ of that of the capsule,
then  laser energy absorption efficiency is $76.2\%$ for both designs.

\begin{table}[htbp]
\caption{Comparisons of 1D implosion performances between the amplifier capsule and central ignition  capsule. Here, $\rho R$ is in g/cm$^2$, $T$  in keV,   $t$   in ns, velocity  in cm/s, mass in mg,   $E_{abs}$ refers to the energy absorbed by capsule, and $\Phi$ refers to the fraction of
burnt fuel.  Convergence ratio is defined as the ratio of initial radius of DT gas to hot spot radius at $t_{ign}$.}
\begin{tabular}{|c|c|c|}
\hline
Capsule & Central & Amplifier\\
\hline
E$_L$ (MJ), P$_L$ (TW) & 1.6, 330 &  9.46, 1080 \\
\hline
E$_{abs}$ (MJ) & 1.19 &  7.24 \\
\hline
mass of DT fuel  & 0.842 &  6.33\\
\hline
mass of CH ablator   & 1.38  &  7.02\\
\hline
$\alpha_{if}$   & 2.9 &  2.37 \\
\hline
implosion velocity  & 4.11 $\times 10^7$ &  3.75 $\times 10^7$ \\
\hline
 Convergence ratio   & 19.8 &  18.6\\
\hline
$t_{fuel}$  & 9.302 &  21.908\\
\hline
$t_{stag}$ & 9.376  &  21.985\\
\hline
$t_{ign}$  & 9.404 &  22.003\\
\hline
 $t_{bang}$ or $t_{primary}$ & 9.457 &  22.025 \\
\hline
$t_{secondary}$  & - &  22.053\\
\hline
@ $t_{fuel}$ ~ $ (\rho R)_{h}$, $T_{i,h}$  & 0.21, 5.3  & 0.28, 6.7  \\
~~~~~~~~~ $ (\rho R)_{c}$, $T_{i,c}$    & 0.65, 0.34  & 1.5, 0.18 \\
\hline
 @ $t_{stag}$ ~ $ (\rho R)_{h}$, $T_{i,h}$  & 0.36, 7.8   & 0.59, 13  \\
~~~~~~~~~ $ (\rho R)_{c}$, $T_{i,c}$  & 0.67, 0.65   & 1.6, 0.34  \\
~~~~~~~~~ $\rho_{c}/\rho_h$, $\xi$    & 14, 0.93  & 28, 3.6 \\
\hline
@ $t_{ign}$ ~ $ (\rho R)_{h}$, $T_{i,h}$ & 0.46, 9.9   & 0.82, 17  \\
~~~~~~~~~ $ (\rho R)_{c}$, $T_{i,c}$   &  0.53, 0.95   & 1.4, 0.45  \\
\hline
@ $t_{bang}$ ~ $ (\rho R)_{f}$, $T_{i,f}$  & 0.82, 17   & - \\
  & whole fuel burnt & \\
\hline
@ $t_{pri}$ ~ $ (\rho R)_{h}$, $T_{i,h}$  & -  & 1.6, 42  \\
~~~~~~~~~ $ (\rho R)_{c}$, $T_{i,c}$   &       & 0.39, 0.36  \\
\hline
@ $t_{sec}$ ~ $ (\rho R)_{f}$, $T_{i,f}$  & -  & 3.2, 85  \\
  & & whole fuel burnt  \\
\hline
 $\Phi$  & 16.2$\%$  &  38.5$\%$\\
\hline
nuclear yield $Y_{id}$ (MJ) & 35.5 &  729 \\
\hline
Target gain G & 22 &  77 \\
\hline
\end{tabular}
\end{table}

\begin{figure}[htbp]
\includegraphics[width=0.49\textwidth]{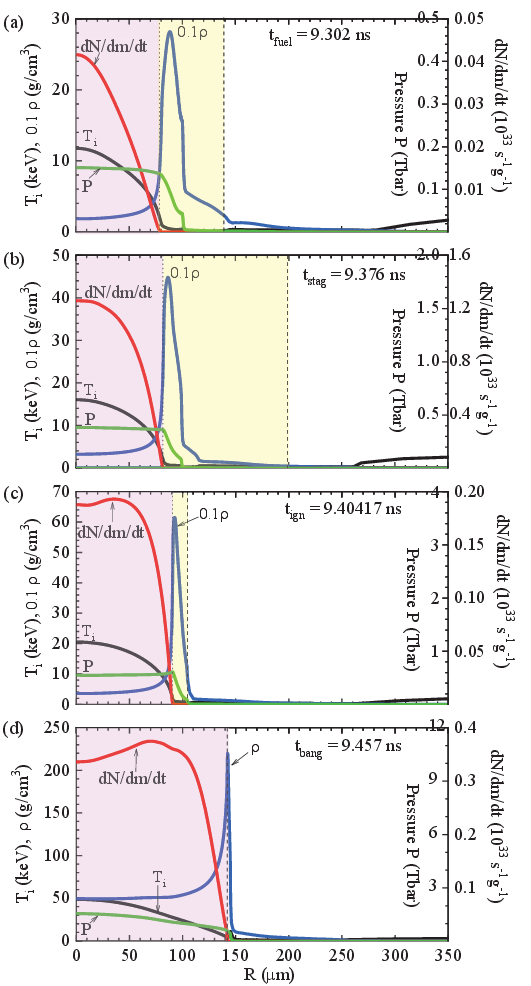}
\caption{(Color online)
Spatial distributions of  $T_{i}$ (black), $\rho$ (blue),  pressure $P$  (green),  and  reaction rate of neutron per unit mass $\frac{dN}{dmdt}$ (red) at $t_{fuel}$ (a), $t_{stag}$  (b), $t_{ign}$ (c), and $t_{bang}$ (d) of the central ignition capsule. The hot spot and cold shell are marked in lavender and yellow, respectively. The  ablator is in white. }
\label{Fig:3}
\end{figure}

\begin{figure}[htbp]
\includegraphics[width=0.49\textwidth]{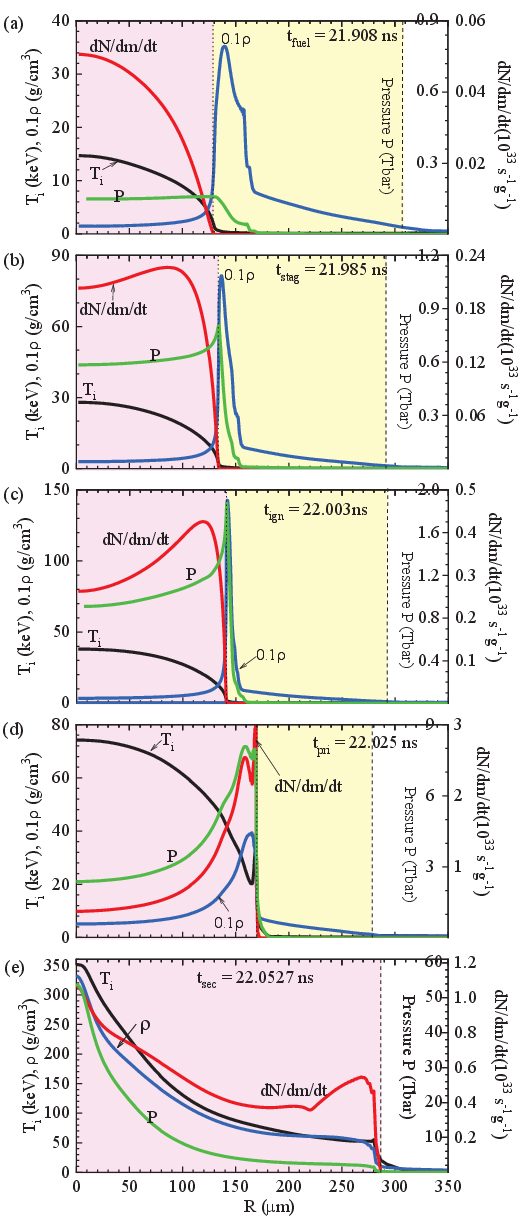}
\caption{(Color online)
Spatial distributions of  $T_{i}$ (black), $\rho$ (blue),  pressure $P$  (green),  and  reaction rate of neutron per unit mass $\frac{dN}{dmdt}$ (red) at $t_{fuel}$ (a), $t_{stag}$  (b), $t_{ign}$ (c), $t_{pri}$ (d), and $t_{sec}$ (e)  of   the amplifier capsule. Same as for Fig. \ref{Fig:3}, the hot spot and cold shell are marked in lavender and yellow, respectively. The  ablator is in white. }
\label{Fig:4}
\end{figure}

Inflight fuel adiabat $\alpha_{if}$ is defined as the ratio of pressure to Fermi-degenerate pressure calculated at the  density maximum at the time of peak velocity.
We have  $\alpha_{if}$ $\sim$ 2.37 for the amplifier  capsule, and  $\alpha_{if}$ $\sim$ 2.9 for the central ignition capsule.
The amplifier capsule reaches its peak implosion velocity  of 3.75 $\times 10^7$ cm/s at 21 ns, and it is   4.11 $\times 10^7$ cm/s at 8.74 ns for the central ignition capsule.
The amplifier design has a slower implosion velocity and a much longer laser pulse.
Table I compares the 1D implosion performances between the two capsules.

For  the central ignition capsule, $t_{fuel}$ = 9.302 ns, $t_{stag}$ = 9.376 ns, $t_{ign}$ = 9.404 ns, and $t_{bang}$ = 9.457 ns, with intervals of 74 ps, 28 ps and 53 ps.
For the amplifier capsule,   $t_{fuel}$ = 21.908 ns, $t_{stag}$ = 21.985 ns, $t_{ign}$ = 22.003 ns, $t_{pri}$ = 22.025 ns, and $t_{sec}$ = 22.053 ns, with intervals of 77 ps, 18 ps, 22 ps and 28 ps.
The central ignition capsule has one explosion within 53 ps after $t_{ign}$, while  the amplifier capsule has two cascading explosions within 50 ps after $t_{ign}$.
For the amplifier capsule, $dN/dt$ reaches its peak   at $t_{pri}$, i.e, $t_{pri}$ = $t_{bang}$ for this design.

For both designs, a notable temperature difference is observed  between ions and electrons in hot spot\cite{Fan2017MRE}. At fuel center, we have $T_i/T_e$ = 118$\%$ at $t_{fuel}$,  106$\%$ at $t_{stag}$,  106$\%$ at $t_{ign}$, and  146$\%$ at $t_{bang}$ for the central ignition capsule; and it is 113$\%$ at $t_{fuel}$,  114$\%$ at $t_{stag}$,  122$\%$ at $t_{ign}$, 158$\%$ at $t_{pri}$, and  500$\%$ at $t_{sec}$ for the amplifier capsule.

The  fusion starts from the central hot spot for both capsules.
We define the place as hot spot boundary where $\frac{dN}{dmdt}$ falls to $1\%$ of its peak.
We use $(\rho R)_h$ and  $T_{i,h}$ to denote the areal density and ion temperature of the hot spot, $(\rho R)_c$ and  $T_{i,c}$   for the cold shell, and $(\rho R)_f$ and  $T_{i,f}$ for the whole fuel, respectively.
The whole fuel is involved in fusion when  $(\rho R)_h$ = $(\rho R)_f$.
From simulations, that happens  at  $t_{bang}$ for the central ignition capsule, while for the amplifier capsule  it is  17 ps later after $t_{pri}$, i.e., 11 ps earlier than $t_{sec}$.

We compare above quantities of the two capsules in Table I.
In Fig. \ref{Fig:3} and  Fig. \ref{Fig:4}, we present  the spatial distributions of  $T_{i}$, $\rho$,   $P$ and $\frac{dN}{dmdt}$ of the two capsules at their characteristic times, respectively.
From Table I,  Fig. \ref{Fig:3}(a) and  Fig. \ref{Fig:4}(a),
it is interesting to note the two capsules have a similar hot spot condition at $t_{fuel}$, with  $(\rho R)_h$ = 0.21 g/cm$^2$ and $T_{i,h}$ = 5.3 keV for the central ignition capsule, and 0.28 g/cm$^2$ and  6.7 keV for the amplifier capsule.
Nevertheless, their cold shells are  very different at this time.
At $t_{fuel}$, the central ignition capsule has a cold shell with  $(\rho R)_c$  = 0.65 g/cm$^2$, $T_{i,c}$ = 0.34 keV  with $\rho_c/\rho_h$ = 15, while it is  1.5 g/cm$^2$ and 0.23 keV  with $\rho_c/\rho_h$ = 25 for the amplifier capsule.
An extremely  dense cold shell with a  large $(\rho R)_c$  and a low $T_{i,c}$   is a key for the amplifier scheme to stop the $\alpha$-particles \cite{Li2019POP}   in later ignition and burn phases.

\begin{figure}[htbp]
\includegraphics[width=0.49\textwidth]{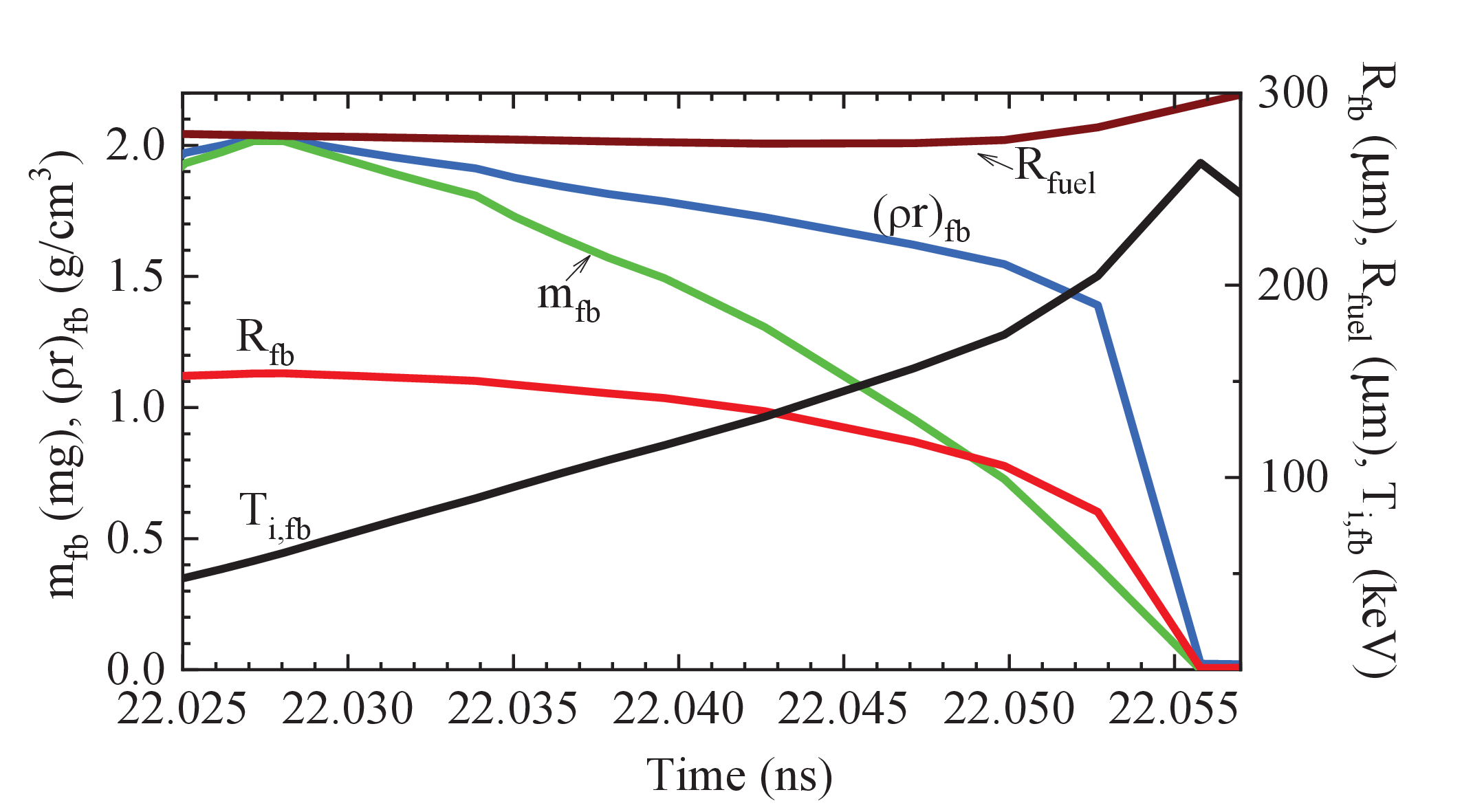}
\caption{(Color online)
Temporal evolutions of $R_{fb}$ (red), $T_{i,fb}$ (black), $m_{fb}$ (green),  $(\rho r)_{fb}$ (blue)  and $R_{fuel}$ (wine) after $t_{pri}$ for  the amplifier capsule. }
\label{Fig:5}
\end{figure}

As shown in Fig. \ref{Fig:3}, the implosion behaviors of the central ignition capsule follow the standard stages of the central ignition scheme. Such as, (1) the central ignition capsule has a constant pressure in hot spot at $t_{fuel}$, $t_{stag}$, $t_{ign}$, and it changes  slowly at $t_{bang}$;  (2) both temperature and fusion reaction rate drop sharply when approaching the shell; and (3) it has only one explosion, and it happens  inside the central hot spot.
However, the amplifier capsule behaves very different.
Though $T_i$ also keeps highest at all times, the amplifier capsule has very different evolutions in $P$ and $\frac{dN}{dmdt}$ because of its extremely dense shell.
Below, we focus on Figs. \ref{Fig:4}(b)-(e) and discuss the characteristics of the amplifier capsule after $t_{fuel}$.
$\newline$ (1) {\it Formation of an extremely compressed shell with a high areal density at stagnation}.
At $t_{stag}$, as shown in Fig. \ref{Fig:4}(a) and Table I,  the shell of the amplifier capsule reaches $\rho_c$ = 810 g/cm$^3$ with $\rho_c$/$\rho_h$  = 28 and $(\rho R)_c$ = 1.6 g/cm$^2$,  even denser  than at $t_{fuel}$.
At this time, $T_i$ reaches its maximum  of  28 keV at center.
However, pressure of the amplifier capsule is not constant anymore.
Both peaks of $P$ and $\frac{dN}{dmdt}$  move from central region toward the shell.
As shown in Fig. \ref{Fig:4}(b), $\frac{dN}{dmdt}$ reaches its peak of  0.22 $\times 10^{33}$ s$^{-1}$ g$^{-1}$ at R $\approx$ 86 $\mu$m  where $T_i$ decreases to 21 keV, and $P$ reaches its peak of  0.81 Tbar at  134 $\mu$m where $T_i$ drops to 2  keV. It indicates that both pressure and fusion rate are dominated by density after $t_{stag}$.
$\newline$ (2) {\it Density dominated ignition and move of fusion peak toward the shell}.
At $t_{ign}$, the  amplfier capsule achieves ignition and releases the amount of fusion energy equal to the input laser energy.
At the center, as shown in Fig. \ref{Fig:4}(c), $T_i$ reaches maximum of 38 keV at center.
In the middle of the shell where $\rho$ peaks,  R is 142 $\mu$m with $\rho_c$ = 1430 g/cm$^3$, $\rho_c$/$\rho_h$  = 42 and $T_i$ = 1.42 keV. At this time, though the shell is still very cold,  both pressure and neutron yield per unit mass are dominated by the extremely high density of shell and their peaks move rapidly toward the middle of shell.
As shown, $\frac{dN}{dmdt}$ reaches its peak of  0.43 $\times 10^{33}$ s$^{-1}$ g$^{-1}$ at 119 $\mu$m with $T_i$ = 20.4 keV, and $P$ reaches its peak of 1.88 Tbar at R $\approx$ 141 $\mu$m with $T_i$ = 1.58 keV.
Obviously, the peak of $P$ moves faster than that of  $\frac{dN}{dmdt}$,
and it almost arrives in the shell middle  where $\rho$ peaks at $t_{ign}$.
$\newline$ (3) {\it Primary explosion in shell and formation of a fireball in the center}.
At $t_{pri}$, $T_i$ reaches maximum of 74 keV with $\rho$ = 51 g/cm$^3$ at center.
It is interesting to see, the peaks of $P$ and $\frac{dN}{dmdt}$ simultaneously arrive in the middle of shell  where $\rho$ peaks, and it leads the primary explosion.
In the shell, $\rho$ reaches its peak of 390 g/cm$^3$ at 164 $\mu$m, with $\rho_c$/$\rho_h$  = 8, $T_i$ = 21 keV,  $P$ = 8 Tbar, and $\frac{dN}{dmdt}$ reaches $3 \times 10^{33}$ s$^{-1}$ g$^{-1}$, as shown in Fig. \ref{Fig:4}(d) and Table I.
Note the shell density becomes lower than at $t_{ign}$ because of the  expansion after ignition.
The primary explosion in the middle of shell violently splits the whole fuel into two parts,  pushing the outer part  to expand while   the inner part to converge spherically to the center. We call this   converging central fuel as   fireball,
and define its boundary as the place where fuel starts to run outwards.
At $t_{pri}$, the fireball mass is 1.93 mg  with a 153 $\mu$m radius,  meaning  30$\%$ of whole fuel is involved in the fireball.
During converging, the fireball becomes smaller and its mass becomes less, because the relatively high pressure of fireball causes the fuel on the  fireball boundary to fly outward.
$\newline$ (4) {\it  Secondary explosion and expansion. }
As converging, all of $\rho$, $T_i$, $P$,  and $\frac{dN}{dmdt}$ in  fireball abruptly increase with their peaks violently moving toward the center.
At $t_{sec}$, 28 ps after $t_{pri}$, the fireball converges at the fuel center and the secondary explosion happens. At this time,  $T_i$, $\rho$,  $P$  and $\frac{dN}{dmdt}$ reach their peaks at the center, with  350 keV,  330 g/cm$^3$,   54 Tbar, and  1.1 $\times 10^{33}$ s$^{-1}$ g$^{-1}$, as shown in  Fig. \ref{Fig:4}(e).
The fireball mass is 0.393 mg, i.e., $\sim$ 6.2$\%$ of whole fuel, with a radius of 82 $\mu$m, averaged density of 169 g/cm$^3$, averaged $\rho R$ of 1.39 g/cm$^2$,
averaged temperature of 204 keV, and  averaged pressure of 19 Tbar.
After $t_{sec}$, the huge pressure at the center violently pushes the whole fuel to expand, and the fireball disappears within 3 ps.
Shown in Fig. \ref{Fig:5} is the temporal evolutions of $R_{fb}$, $m_{fb}$,  $T_{i,fb}$  and $(\rho r)_{fb}$ after $t_{pri}$, where $R_{fb}$ is   radius,  $m_{fb}$ is  mass, $T_{i,fb}$ is  averaged temperature,  and $(\rho r)_{fb}$ is averaged areal density of the fireball.
The radius of whole fuel $R_{fuel}$ is also presented for comparison.  Hence, the fireball can last about 30 ps after $t_{pri}$ within a radius of about 150 $\mu$m, with temperature increasing as time and peaking at  260 keV.

\begin{figure}[htbp]
\includegraphics[width=0.49\textwidth]{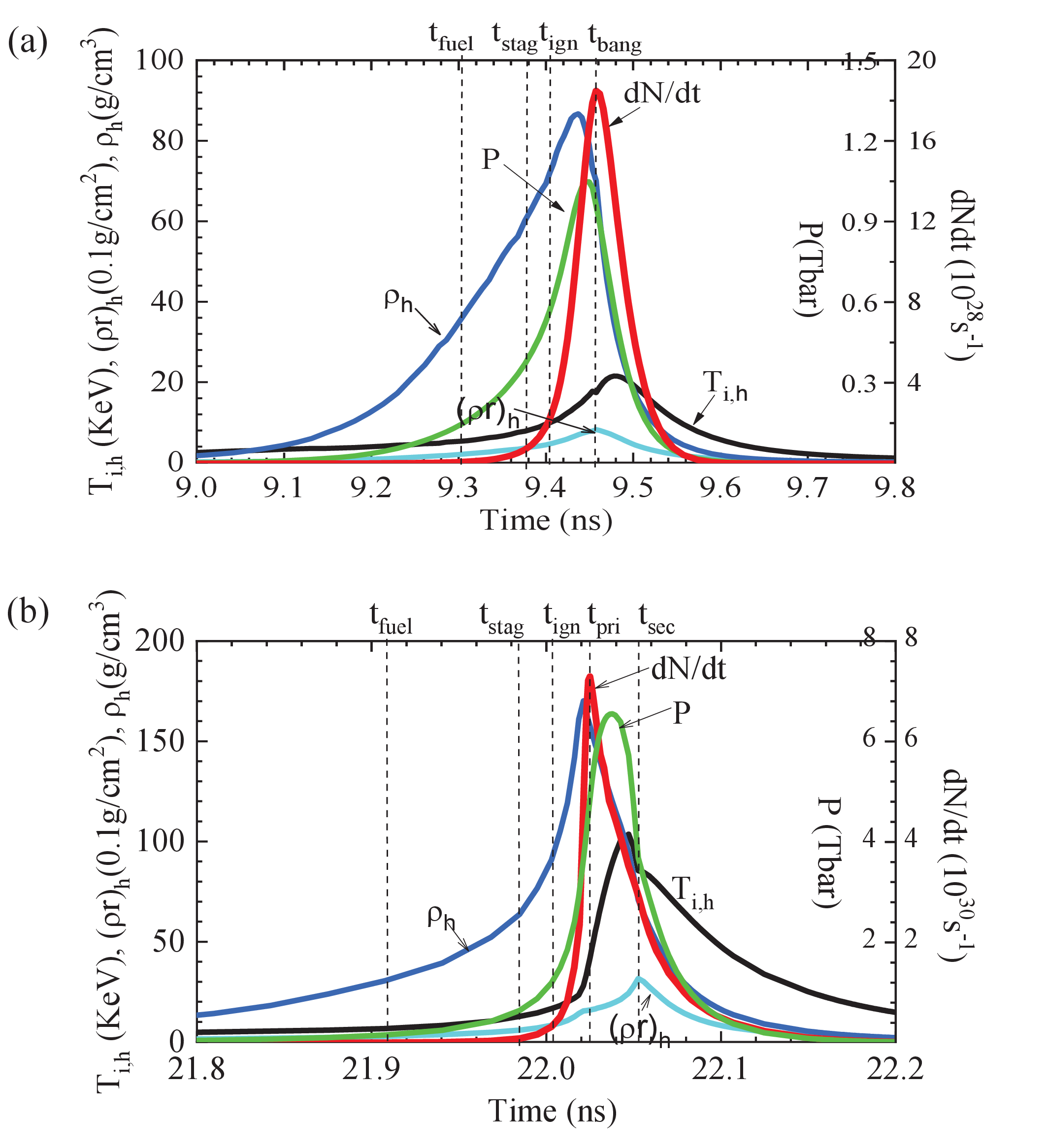}
\caption{(Color online)
Temporal evolutions of $T_{i,H}$ (black), $\rho_H$ (cyan), $(\rho r)_H$ (blue),   $P$  (green) and $\frac{dN}{dt}$ (red)  for the central ignition capsule (a) and amplifier capsule (b). Characteristic  times  are labelled with black dashed lines.}
\label{Fig:6}
\end{figure}

\begin{figure}[htbp]
\includegraphics[width=0.45\textwidth]{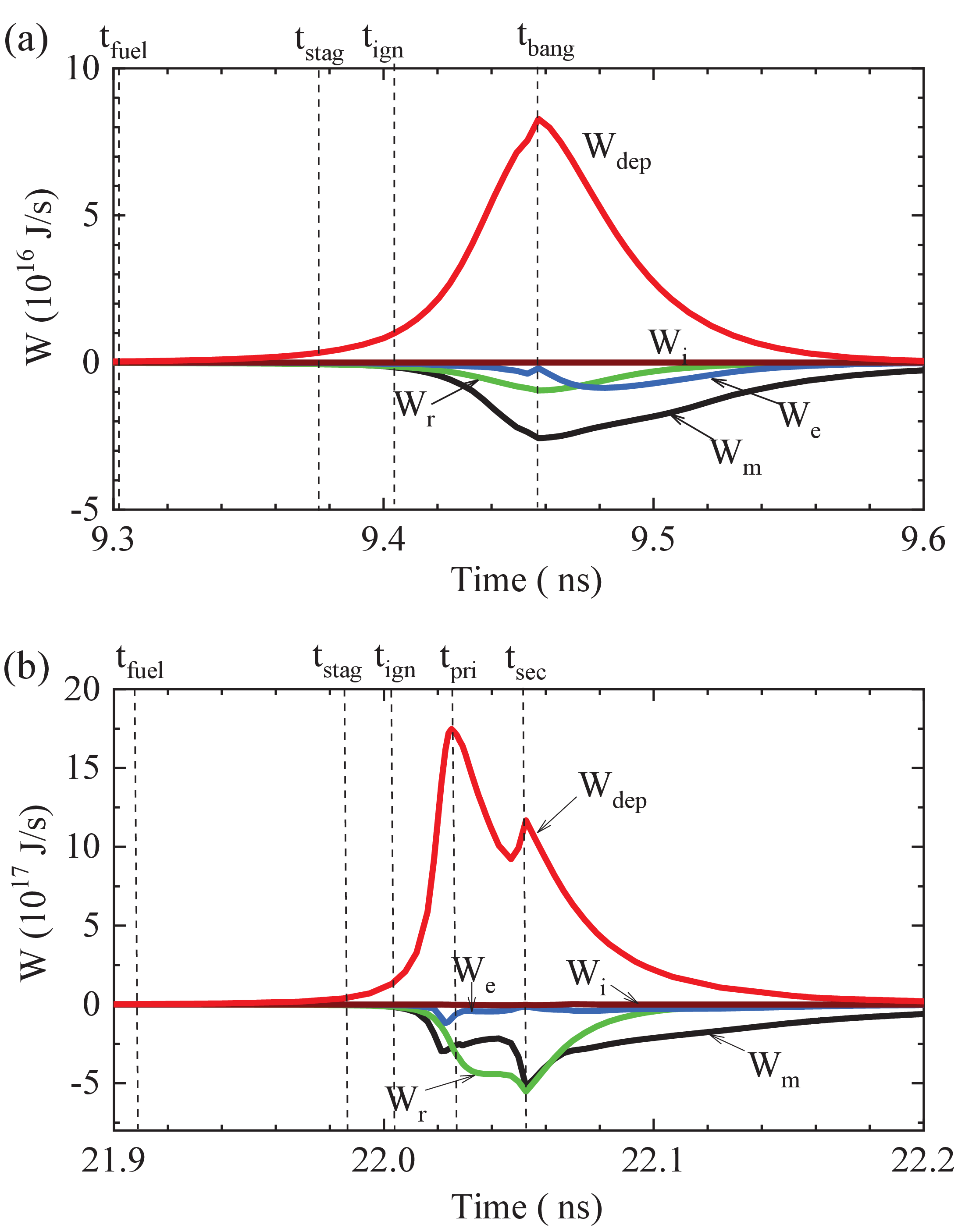}
\caption{(Color online)
Temporal evolutions of $W_{dep}$ (red),  $W_m$ (black),     $W_r$ (green),    $W_e$ (blue), and  $W_i$ (wine) of hot spot for the central ignition capsule (a) and amplifier capsule (b). Times of $t_{stag}$, $t_{pri}$ and $t_{sec}$ are labelled with black dashed lines.}
\label{Fig:7}
\end{figure}

\begin{figure}[htbp]
\includegraphics[width=0.49\textwidth]{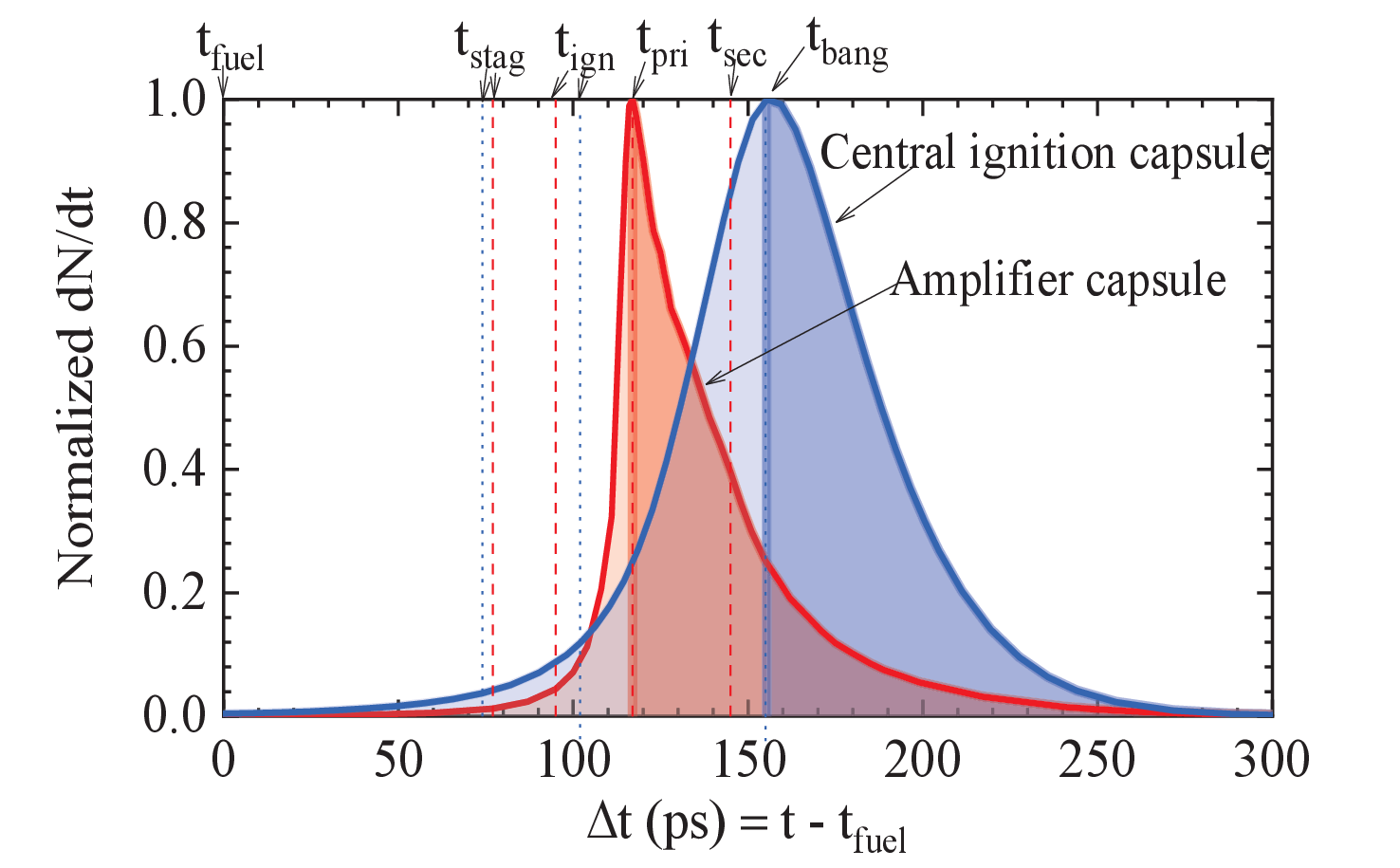}
\caption{(Color online)
Temporal evolutions of normalized $\frac{dN}{dt}$ for the central ignition capsule (blue) and the amplifier capsule (red). Characteristic  times  are labelled with blue dotted lines for  the central ignition capsule and with red dashed lines for the amplifier capsule.}
\label{Fig:8}
\end{figure}

\begin{figure}[htbp]
\includegraphics[width=0.45\textwidth]{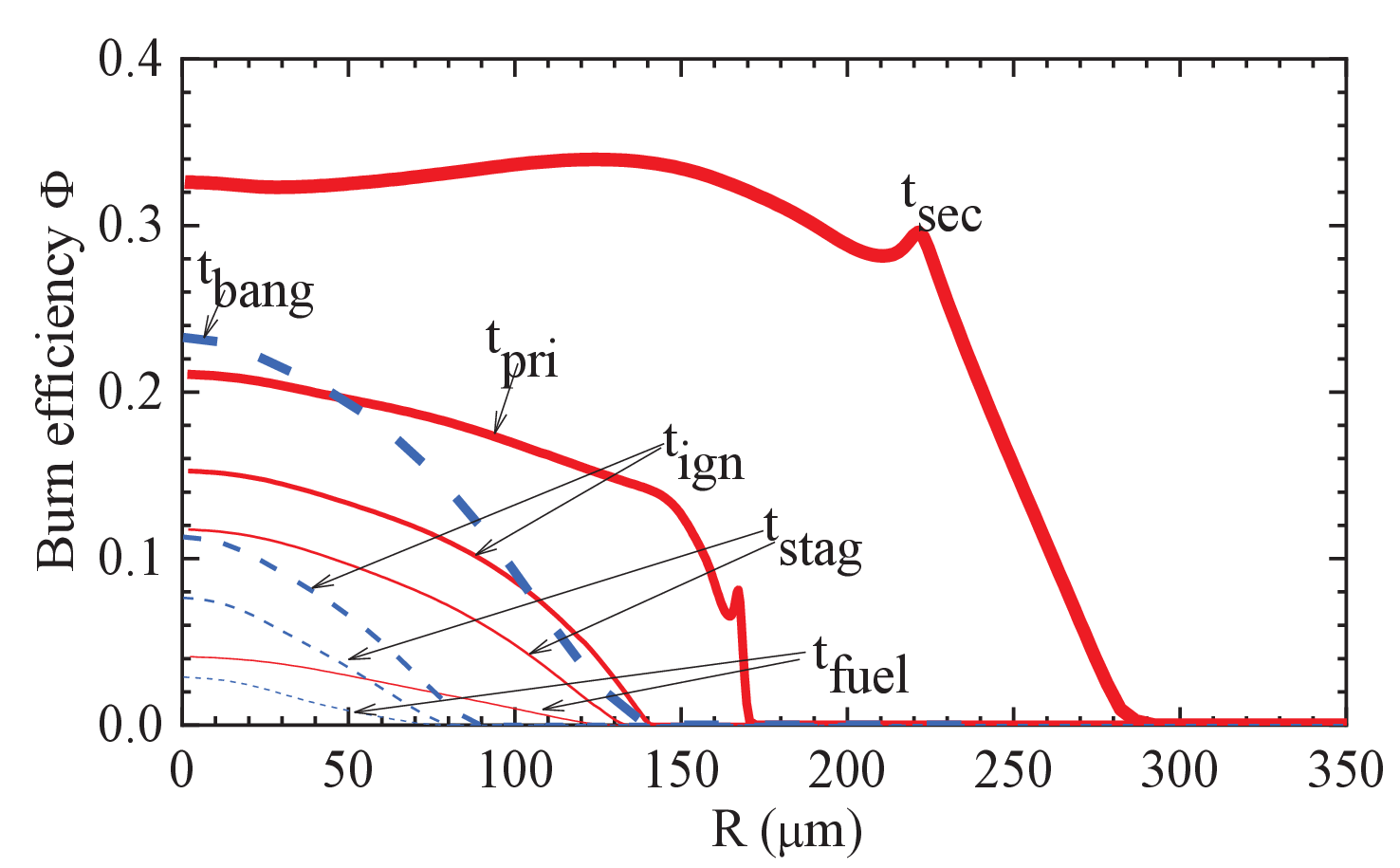}
\caption{(Color online)
Spatial distribution  of the fraction of burnt fuel $\Phi$ at the characteristic times of the central ignition capsule (blue dashed lines) and the amplifier capsule (red solid lines).}
\label{Fig:9}
\end{figure}

\begin{figure*}[htbp]
\includegraphics[width=1.0\textwidth]{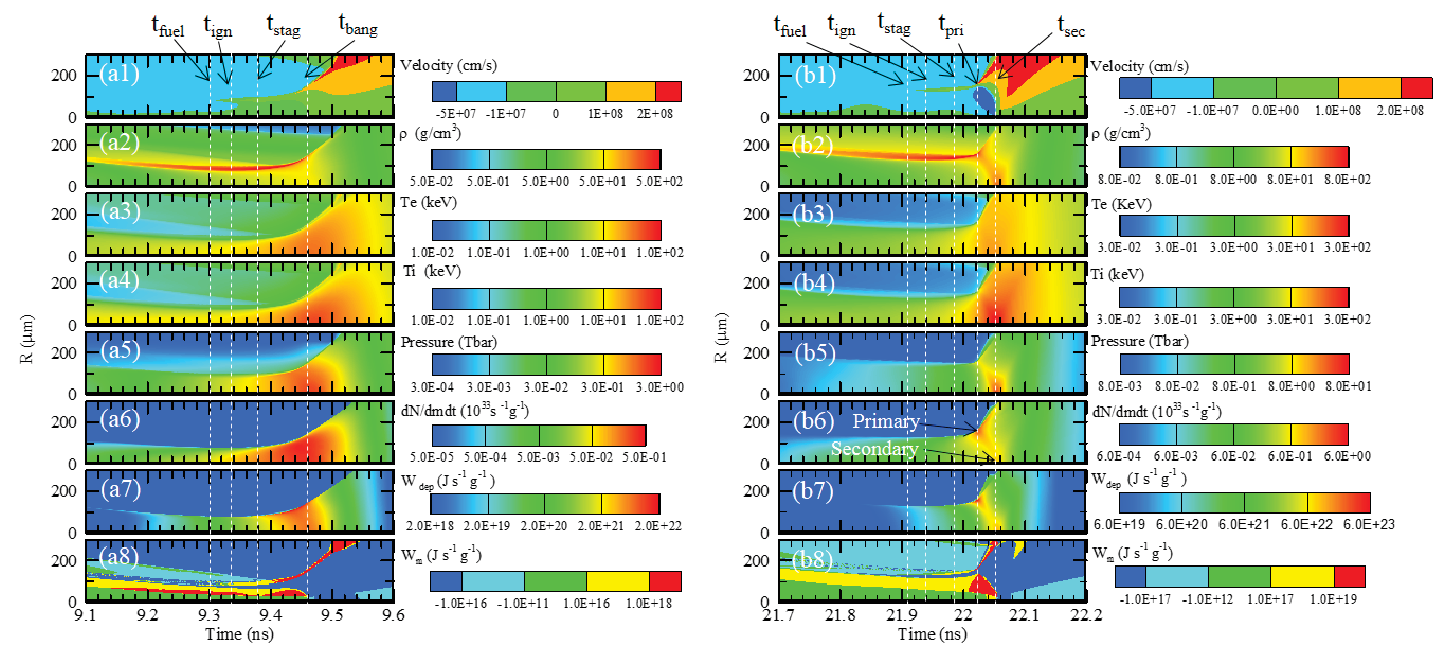}
\caption{(Color online)
Radial distribution and temporal evolution of  $v$ (a1, b1),   $\rho$ (a2, b2),  $T_e$ (a3, b3),   $T_i$ (a4, b4),  $P$  (a5, b5),   $\frac{dN}{dmdt}$ (a6, b6),  $W_{dep}$(a7, b7), and $W_m$ (a8, b8)  for the central ignition capsule  (left) and the amplifier capsule (right). Characteristic times  are labelled with white dashed lines. }
\label{Fig:10}
\end{figure*}

\begin{figure*}[htbp]
\includegraphics[width=1.0\textwidth]{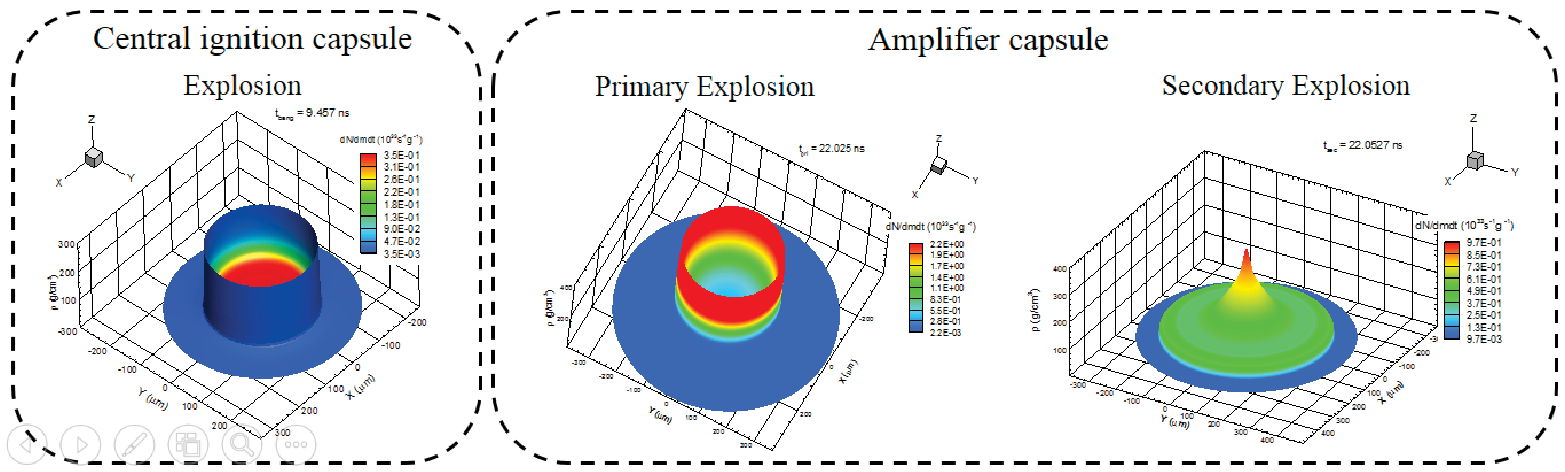}
\caption{(Color online)
Spacial distributions of $\rho$ (shown by z-axis) and $\frac{dN}{dmdt}$ (shown by color) in x-y plane at $t_{bang}$ of the central ignition capsule (left) and $t_{pri}$ and $t_{sec}$ of the amplifier capsule (right). }
\label{Fig:11}
\end{figure*}

It is worth discussing an interesting  phenomenon  in Fig. \ref{Fig:4} (d), in which the burn propagation is seen to be ``trapped'' inside the sharp boundary located at the outer shell surface. This is because the rate of $\alpha$-particle local energy deposition exceeding to thermal heat flux at this place.  From Ref. \citenum{MTV}, we have $W_{dep}$ $\sim$ $T^2$ and $W_e$ $\sim$ $T^{\frac{7}{2}}/\rho R^2$. The explosion can happen at the outer shell surface at $t_{pri}$ when $W_{dep}/W_e > $  1 in case it is  $W_e$ which dominates the cooling of the fuel at the outer shell surface.
Considering $\rho_c/\rho_h$ is  a crucial factor for the amplifier scheme,  we can define $\rho_c/\rho_h$ and  $\xi$ = $a (\rho R)_c /T_{i,c}^{\frac{3}{4}}$ at stagnation as the trigger criterions for the first explosion in shell. Here, $a$  is a constant to be determined by simulations.
Here, we take  $a$ =1, with $(\rho R)_c$ in g/cm$^2$ and $T_{i,c}$ in keV.
At $t_{stag}$, we have $\rho_c/\rho_h$ = 14 and $\xi$ = 0.93  for the central ignition capsule,
and $\rho_c/\rho_h$ = 28 and $\xi$ = 3.6 for the amplifier capsule, as shown in Table. Hence, both $\xi$ and $\rho_c/\rho_h$ of the  amplifier capsule are obviously larger than that of the central ignition capsule.

Shown in Fig. \ref{Fig:6} is the temporal evolutions of $T_{i,H}$, $\rho_H$, $(\rho r)_H$,   $P$, and $\frac{dN}{dt}$  for the two capsules.
For the central ignition capsule, as shown,   $\rho_h$, $(\rho r)_h$ and $P$ almost reach their peaks around $t_{bang}$ when $\frac{dN}{dt}$  reach its peak, while $T_{i,h}$ reaches its peak a little later than  $t_{bang}$ because of the strong heat aroused  by  explosion.
In contrast, for the amplifier capsule, $\rho_H$  reaches  its   peak at $t_{pri}$ while  $(\rho r)_H$ is still increasing;
and both $(\rho r)_H$  and $T_{i,H}$ reach their peaks around $t_{sec}$ while $\rho_H$   has already dropped to half of its peak.
Obviously, the primary explosion of  the amplifier capsule is dominated by density,  while the secondary explosion is dominated by areal density and temperature.

Presented in Fig. \ref{Fig:7} is the  temporal evolutions of $W_{dep}$, $W_m$, $W_r$, $W_e$ and  $W_i$ of hot spot  for the two capsules.
As shown, between $t_{stag}$ and $t_{ign}$, $E$ is mainly raised by  $W_{dep}$ while lost by  $W_m$ for both capsules.
However, it is a little different after explosion happens.
For the central ignition capsule,
All $W_{dep}$, $|W_m|$ and  $|W_r|$  reach their  peaks at  $t_{bang}$ when   explosion happens, indicating that whole fuel expands immediately after  explosion.
In contrast, for the amplifier capsule,   $W_{dep}$ has two peaks at $t_{pri}$ and $t_{sec}$, respectively, due to two cascading explosions.
In addition, $|W_r|$ of the amplifier capsule greatly increases after $t_{pri}$  and reaches its maximum at $t_{sec}$; and $|W_m|$ reaches its 1st peak a short time earlier before  $t_{pri}$ and   2nd peak at $t_{sec}$, due to the strong expansions caused by the two explosions.
The contributions of $W_i$ can be completely neglected for both capsules.

It is interesting to compare  the  temporal  evolutions of the fusion rate $\frac{dN}{dt}$ between the two capsules in Fig. \ref{Fig:8}.
As shown, the central ignition capsule has almost the same rising and falling rates of $\frac{dN}{dt}$ around bang time. In contrast, the amplifier capsule has a much slower falling part, due to the contribution of   secondary explosion.
As a  comparison, the yield released by the amplifier capsule  after  $t_{bang}$ is 4.8 times that before, while it is 1.25 times for the  central ignition capsule. It demonstrates that the amplifier capsule releases  a significant additional  yield in burn stage after ignition.
Presented in Fig.\ref{Fig:9} is spatial distributions of  the fraction of
burnt fuel $\Phi$ at the characteristic times of the two capsules. As shown,  $\Phi$ at the center of the central ignition capsule at $t_{bang}$ is even higher that of the amplifier capsule at $t_{pri}$, but $\Phi$ of  the central ignition capsule decreases faster than the amplifier capsule along the radial direction. It is interesting to note that $\Phi$ of the amplifier capsule increases remarkably with a very fast burn propagation from $t_{pri}$ to $t_{sec}$.
As a result, we have $\Phi$ = 38.5 $\%$ with $Y_{id}$ = 729 MJ and G = 77 for the amplifier capsule, while $\Phi$ = 16.2 $\%$ with $Y_{id}$ = 35.5 MJ and G = 22 for the central ignition capsule.

To have an overall picture on differences between the amplifier capsule and the central ignition capsule, we present in Fig. \ref{Fig:10}  the simulated evolutions in space  and time of fluid velocity $v$,  $\rho$,  $T_e$,   $T_i$,  $P$,    $\frac{dN}{dm dt}$,   $W_{dep}$  and  $W_m$ for  the two capsules within a time window before stagnation and after explosions.
For the central ignition capsule, its stages of (1) ablation and implosion can be   seen in Frames a1 and a8,  (2) stagnation and formation of a hot spot can be seen in Frames a1 to a8, (3) ignition in hot spot can be seen in Frames a3 to a7, (4) burn and explosion  can be seen in Frames a1 to a8.
For the amplifier capsule, its stages of (1) ablation and implosion can be  seen in Frames b1 and b8,
(2) formation of an extremely compressed shell can be  seen from Frames b2,
(3) density dominated ignition and move of fusion peak toward the shell can seen from  Frames b2 and b7,
(4) primary explosion in shell and formation of a fireball in the center can be seen in Frames b1 to b8;
(5) secondary explosion in the extremely hot and dense fuel can be seen in Frames b1 to b8.
Comparing Frames b3 and b4 with a3 and a4, we can see that temperature in the amplifier capsule increase much more violently during explosions than in the central ignition capsule.

Moreover, we present in Fig. \ref{Fig:11} the spacial distributions of $\rho$ and $\frac{dN}{dm dt}$ at $t_{bang}$ for the central ignition capsule and at $t_{pri}$ and $t_{sec}$ for the amplifier capsule. As shown, the two capsules have quite different characteristics in the spacial distribution of $\rho$  and $\frac{dN}{dm dt}$  when their explosions happen.
For the central ignition capsule, its explosion is dominated by temperature and happens inside the central hot spot at a relatively low density, which massive fuel mass is wasted in the cold shell of high density.
In contrast, the primary explosion of the amplifier capsule is dominated by density and  happens inside the cold shell, leading to almost half of the fuel involved in the explosion.
Indeed, the secondary explosion of  the amplifier capsule happens in the central region, but with all of density, temperature and pressure  reaching their peaks at the center at this time and thus leading to a very efficient burn-up, as shown in Fig. \ref{Fig:9}. This is quite different from the central ignition scheme which density peaks in the cold shell when its explosion happens.

Finally, it is worth comparing our amplifier scheme with the shock ignition scheme \cite{Betti2007PRL} in the following.
$\newline$ (1) The shock ignition scheme is not the conventional inertial confinement fusion and needs an ignitor shock to heat its central hot spot to ignite the assembled fuel. In contrast, the amplifier scheme is realized fully under inertial confinement, with no need to add any ignitor shock.
$\newline$ (2) The resulting fuel assembly of the shock ignition features a hot-spot pressure greater than the surrounding dense fuel pressure. In contrast,  the amplifier scheme has a surrounding dense fuel pressure greater than the hotspot pressure at and after stagnation.
$\newline$ (3) The shock ignition needs to launch a spherically converging shock in the latest stage of the implosion to attain a peaked pressure. In contrast, the amplifier scheme does not need to launch such a converging shock in the latest stage of the implosion. Its peak pressure in shell for primary explosion is generated by inertial confinement, and its peak pressure at core for secondary explosion is
generated by the primary explosion.
$\newline$ (4) For the shock ignition, in order to maximize the hotspot peak pressure of the final assembly, the ignitor shock must collide with the return shock inside the dense shell and near its inner surface. In contrast, the amplifier scheme does not  have  such requirement.
$\newline$ (5) For the shock ignition, its two new shocks are generated as a result of the collision: an inward and outward moving shock. In contrast, the inward and outward moving shocks of the amplifier scheme is a result of the primary explosion, not a result of collision.
$\newline$ (6) For the shock ignition, launching a shock during the final stage of the implosion and timing it with the return shock is to generate a nonisobaric assembly.
In contrast, the nonisobaric of the amplifier scheme is leaded by the density dominated nuclear reaction in shell.
$\newline$ (7) For the shock ignition, because of the relatively high laser intensity $6 \times 10^{15}$ W/cm$^2$ in the spike, a significant amount of hot electrons can be generated by the laser plasma instabilities.
In contrast, with no need of such high laser intensity, the amplifier scheme does not have such high laser intensity aroused hot electron issues.
$\newline$ (8) 
With no need of the ignitor shock, the amplifier scheme does not have the ignitor shock aroused  hydrodynamic instabilities at the end of the acceleration phase.
$\newline$ (9) The shock ignition is similarly to fast and impact ignition, which shock ignition is induced separately from the compression.  In contrast, with no need to add ignitor shock separately, the primary explosion of the amplifier scheme happens automatically inside the extremely compressed shell under inertial confinement.

In summary,  we   propose  an amplifier scheme to produce additional gain via cascading explosions,  which can be realized either by direct-drive or by indirect-drive.
In novel scheme, the primary explosion is dominated by density and can be generated at a low convergence ratio, and the secondary explosion is driven directly by the primary explosion.
To compare the differences between this new amplifier scheme and the central ignition scheme, we presented the simulations results on a  direct-drive amplifier capsule with 6.33 mg DT fuel under  a  9.46 MJ laser and a central ignition  capsule with 0.842 mg DT fuel under a 1.6 MJ laser.
It may seem not a fair comparison because  the amplifier capsule uses 5.9 times laser energy of the central ignition capsule, but note it drives the 7.5 times fuel mass.
As a result, the yield released by the amplifier capsule  after bang time  is 4.8 times that before, while it is 1.25 times for the  central ignition capsule. This demonstrates that the amplifier capsule can release remarkable additional  yield  in burn stage after ignition.
The amplifier scheme can greatly relax the $\rho R T$ hot spot condition and  the stringent requirements on target design, target fabrication, and laser engineering issues to achieve a high gain.
According to our studies, both $\rho_c/\rho_h$ and $\xi$ = $a (\rho R)_c /T_{i,c}^{\frac{3}{4}}$  can be considered as the trigger criterions for the first explosion in shell, and we are doing the parameter scan to identify them by simulations.
We will  optimize the amplifier design under a lower laser energy, which  is a key for  the commercial feasibility of fusion power station.
More interesting physics and designs are to be explored in this novel scheme.
The  extremely hot and dense fireball generated in the amplifier scheme provides a room for  novel target designs towards clean fusion energy.
Nevertheless,  the requirement for a high density ratio of the cold shell to the hot spot in the amplifier capsule may be challenging and lead to a hydrodynamic unstable design.
We will investigate the beam-to-beam overlapping non-uniformity and hydrodynamic instabilities on the amplifier capsule in our future works.

{\bf ACKNOWLEDGMENTS}
K. L.  appreciates Professor Vladimir Tikhonchuk of the ELI-Beamlines for beneficial discussions on our novel scheme and appreciate S. Atzeni and J. Meyer-ter-Vehn for their very nice book, Ref. \citenum{MTV} in helping to understand and describe the novel phenomena.
This work is supported by the National Natural Science Foundation of China (Grant No. 12035002).

\end{CJK*}


\begin{thebibliography}{00}

\bibitem{MTV} S. Atzeni and J. Meyer-ter-Vehn, {\it The Physics of Inertial Fusion: Beam Plasma Interaction, Hydrodynamics, Dense Plasma Physics} (Clarendon Press, Oxford, 2004); ``Report of the Fusion Energy Sciences Workshop on Inertial Fusion Energy'', U. S. Department of Energy, (2023).
\bibitem{IFEneedsreport} https://science.osti.gov/-/media/fes/pdf/workshop-reports/2023/IFE-Basic-Research-Needs-Final-Report.pdf

\bibitem{Abu-Shawareb2024PRL} H. Abu-Shawareb, R. Acree, P. Adams, J. Adams, B. Addis, R. Aden, P. Adrian, B. B. Afeyan,  M. Aggleton, L. Aghaian  {\it et al.}, ``Achievement of Target Gain Larger than Unity in an Inertial Fusion Experiment'', {\it Phys. Rev. Lett.} {\bf 132}, 065102 (2024).

\bibitem{Hurricane2024PRL} O. A. Hurricane, D. A. Callahan, D. T. Casey, A. R. Christopherson, A. L. Kritcher, O. L. Landen,  S. A. Maclaren, R. Nora, P. K. Patel, J. Ralph {\it et al.}, ``Energy Principles of Scientific Breakeven in an Inertial Fusion Experiment'', {\it Phys. Rev. Lett.} {\bf 132}, 065103 (2024).

\bibitem{Rubery2024PRL} M. S. Rubery, M. D. Rosen, N. Aybar, O. L. Landen, L. Divol, C. V. Young,
  C. Weber, J. Hammer, J. D. Moody, A. S. Moore   {\it et al.}, ``Hohlraum Reheating from Burning NIF Implosions'',  {\it Phys. Rev. Lett.} {\bf 132}, 065104 (2024).
\bibitem{Pak2024PRE} A. Pak, A. B. Zylstra, K. L. Baker, D. T. Casey,  E. Dewald, L. Divol,
   M. Hohenberger, A. S. Moore, J. E. Ralph, D. J. Schlossberg  {\it et al.}, ``Observations and properties of the first laboratory fusion experiment to exceed a target gain of unity,'' {\it Phys. Rev. E} {\bf 109}, 025203 (2024).
\bibitem{Kritcher2024PRE} A. L. Kritcher, A. B. Zylstra, R. Weber, O. A. Hurricane, D. A. Callahan, D. S. Clark, L. Divol, D. E. Hinkel, K. Humbird, O. Jones   {\it et al.}, ``Design of the first fusion experiment to achieve target energy gain G $>$ 1,''  {\it Phys. Rev. E} {\bf 109}, 025204 (2024).

\bibitem{5.2MJ}	See https://www.energy.gov/cfo/articles/fy-2025-budget-justification for more information about the new fusion yield record of 5.2 MJ on the NIF.

\bibitem{Lan2014POP} K. Lan, J. Liu, D. Lai, W. Zheng, and X. He, ``High flux symmetry of the spherical hohlraum with octahedral 6LEHs at the hohlraumto-capsule radius ratio of 5.14'', {\it Phys. Plasmas} {\bf 21}, 010704 (2014).
\bibitem{Divol2017POP}  L. Divol, A. Pak, L. F. Berzak Hopkins, ``Symmetry control of an indirectly driven high-density-carbon implosion at high convergence and high velocity,'' {\it Phys. Plasmas} {\bf 24}, 056309 (2017).
\bibitem{Craxton2020DPP} S. Craxton, ``A new beam configuration to support both spherical hohlraums and symmetric direct drive''.  The 62nd Annual Meeting of the American Physical Society Division of Plasma Physics,  November 9-13, 2020 in U.S.A.
\bibitem{Lan2021PRL} K. Lan, Y. Dong, J. Wu, Z. Li, Y. Chen, H. Cao,  L. Hao, S. Li, G. Ren, W. Jiang {\it et al.}, ``First inertial confinement fusion implosion experiment in octahedral spherical hohlraum,'' {\it Phys. Rev. Lett.} {\bf 127}, 245001(2021).
\bibitem{Lan2022MRE} K. Lan, ``Dream fusion in octahedral spherical hohlraum,'' {\it Matter Radiat. Extremes} {\bf 7}, 055701 (2022).
\bibitem{Ralph2024NP} J. E. Ralph, J. S. Ross, A. B. Zylstra, A. L. Kritcher, H. F. Robey, C. V. Young, O. A. Hurricane, A. Pak, D. A.Callahan, K. L. Baker {\it et al.}, ``The impact of low-mode symmetry on inertial fusion energy output in the burning plasma state,'' {\it Nature Communications} {\bf 15}, 2975 (2024).

\bibitem{Strozzi2017PRL} D. J. Strozzi, D. S. Bailey, P. Michel, L. Divol, S. M. Sepke, G. D. Kerbel, C. A. Thomas, J. E. Ralph, J. D. Moody, and M. B. Schneider, ``Interplay of Laser-Plasma Interactions and Inertial Fusion Hydrodynamics,'' {Phys. Rev. Lett.} {\bf 118}, 025002 (2017).
\bibitem{Tikhonchuk2021MRE} V. T. Tikhonchuk, T. Gong, N. Jourdain, O. Renner, F. P. Condamine, K. Q. Pan, W. Nazarov, L. Hudec, J. Limpouch, R. Liska {\it et al.} Studies of laser-plasma interaction physics with low-density targets for direct-drive inertial confinement fusion on the Shenguang III prototype. {\it Matter Radiat. Extremes} {\bf 6}, 025902 (2021).
\bibitem{incoherent2023MRE} Y. Guo, X. Zhang, D. Xu, X. Guo, B. Shen, and K. Lan, ``Suppression of stimulated Raman scattering by angularly incoherent light, towards a laser system of incoherence in all dimensions of time, space, and angle,'' {\it Matter Radiat. Extremes} {\bf 8}, 035902 (2023).

\bibitem{Goncharov2000POP}  V. N. Goncharov, S. Skupsky, T. R. Boehly, J. P. Knauer, P. McKenty, V. A. Smalyuk, R. P. J. Town, O. V. Gotchev, R. Betti, and D. D. Meyerhofer, ``A model of laser imprinting,'' {\it Phys. Plasmas} {\bf 7}, 2062 (2000).
\bibitem{Clark2018POP}  D. S. Clark, A. L. Kritcher, S. A. Yi, A. B. Zylstra, S. W. Haan, and C. R. Weber, ``Capsule physics comparison of National Ignition Facility implosion designs using plastic, high density carbon, and beryllium ablators,'' {\it Phys. Plasmas} {\bf 25}, 032703 (2018).
\bibitem{Qiao2021PRL} X. Qiao and K. Lan, ``Novel Target Designs to Mitigate Hydrodynamic Instabilities Growth in Inertial Confinement Fusion,'' {\it Phys. Rev. Lett.} {\bf 126}, 185001(2021).
\bibitem{Do2022PRL} A. Do,  C. R. Weber, E. L. Dewald, D. T. Casey, D. S. Clark, S. F. Khan, O. L. Landen, A. G. MacPhee, and V. A. Smalyuk, ``Direct measurement of ice-ablator interface motion for instability   mitigation in indirect drive ICF implosions,'' {\it Phys. Rev. Lett.} {\bf 129}, 215003 (2022).

\bibitem{10MJ}	Z. Sui and K. Lan, ``Driver at 10MJ and 1 shot/30 min for inertial confinement fusion at high gain: Efficient, compact, low-cost, low laser-plasma instabilities, beam color selectable from 2$\omega$/3$\omega$/4$\omega$, applicable to multiple laser fusion schemes,'' {\it Matter Radiat. Extremes} {\bf 9}, 043002(2024).


\bibitem{Nuckolls1972Nature} J. Nuckolls, L. Wood, A. Thiessen, and G. Zimmerman, ``Laser compression of
matter to super-high densities: Thermonuclear (CTR) applications,'' {\it Nature} {\bf 239},
139 (1972).
\bibitem{Campbell2017MRE} E. M. Campbell, V. N. Goncharov, T. C. Sangster, S. P. Regan, P. B. Radha, R. Betti, J.F. Myatt, D.H. Froula, M.J. Rosenberg, I.V. Igumenshchev {\it et al.}, ``Laser-direct-drive program: Promise, challenge, and path forward,'' {\it Matter Radiat. Extremes} {\bf 2}, 37 (2017).
\bibitem{Gopalaswamy2019Nature} V. Gopalaswamy, R. Betti, J. P. Knauer, N. Luciani, D. Patel, K. M. Woo, A. Bose, I. V. Igumenshchev, E. M. Campbell, K. S. Anderson {\it et al.}, ``Tripled yield in direct-drive laser fusion through statistical modelling,'' {\it Nature} {\bf 565}, 581 (2019).

\bibitem{LYS2024POP}  Y. Li, K. Lan, H. Cao, Y.-H. Chen, X. Zhao, and Z. Sui, ``Amplifier scheme: increasing burn efficiency via cascading explosions for inertial confinement fusion,'' to be published in {\it Phys. Plasmas}.

\bibitem{Lindl2018POP} J. Lindl, S. W. Haan, O. L. Landen, A. R. Christopherson, and R. Betti, ``Progress toward a self-consistent set of 1D ignition capsule metrics in ICF,'' {\it Phys. Plasmas} {\bf 25}, 122704 (2018).

\bibitem{Feng1999} T. Feng, D. Lai, and Y. Xu, ``An artificial-scattering iteration method for
calculating multi-group radiation transfer problem,'' {\it Chin. J. Comput. Phys.} {\bf 16}, 199 (1999).
\bibitem{Lan2017POP} K. Lan, X. Qiao, P. Song, W. Zheng, B. Qing, and J. Zhang, ``Study on laser-irradiated Au plasmas by detailed configuration accounting atomic physics,'' {\it Phys. Plasmas} {\it 24},102706 (2017).
\bibitem{Qiao2019PPCF} X. Qiao and K. Lan, ``Study of high-Z coated ignition target by DCA atomic physics for direct-drive ICF,'' {\it Plasma Phys. Control. Fusion} {\bf 61}, 014006 (2019).

\bibitem{Fan2017MRE} Z. Fan, Y. Liu, B. Liu, C. Yu, K. Lan, and J. Liu, ``Non-equilibrium between ions and electrons inside hot spots from National Ignition Facility experiments,'' {\it Matter  Radiat.  Extremes} {\bf 2},3 (2017).

\bibitem{Li2019POP} K. Li, and K. Lan, ``Escape of $\alpha$-particle from hot-spot for inertial confinement fusion,'' {\it Phys. Plasmas} {\bf 26}, 122701 (2019).





\bibitem{Betti2007PRL} R. Betti, C. D. Zhou, K. S. Anderson, L. J. Perkins, W. Theobald, and A. A. Solodov, ``Shock Ignition of Thermonuclear Fuel with High Areal Density,'' {\it Phys. Rev. Lett.} {\bf 98}, 155001 (2007).

\end{thebibliography}
\end{document}